\documentstyle[prd,aps,eqsecnum,12pt]{revtex}

\title{
Lagrangian description of fluid flow with pressure in 
relativistic cosmology}

\author{
Hideki Asada\footnote{Electronic address:
asada@phys.hirosaki-u.ac.jp}}
\address{
Faculty of Science and Technology, Hirosaki University, Hirosaki
036-8561, Japan\\
Max-Planck-Institut f\"ur Astrophysik, 
Karl-Schwarzschild-Str. 1, D-85741 Garching, Germany
} 

\begin{document}
\maketitle

\begin{abstract}
The Lagrangian description of fluid flow in relativistic cosmology 
is extended to the case of flow accelerated by pressure. 
In the description, the entropy and the vorticity are obtained exactly 
for the barotropic equation of state. 
In order to determine the metric, the Einstein equation is solved 
perturbatively, when metric fluctuations are small but 
entropy inhomogeneities are large. 
Thus, the present formalism is applicable to the case when 
the inhomogeneities are small in the large scale but locally nonlinear. 
\end{abstract}

\begin{flushleft}
PACS Number(s); 98.80.Hw, 04.25.Nx 
\end{flushleft}

\section{Introduction}
Understanding the evolution of fluids in the expanding universe is 
of great interest in the cosmology. 
Here, we develop the relativistic version of the Lagrangian
description of fluid flow for the following reason: 
First, let us make a comparison between the Eulerian description 
and the Lagrangian one. 
In the Eulerian approach, all variables are expanded in series. 
Consequently, the density contrast should be small. 
On the other hand, the Lagrangian framework is based 
on the description along the fluid flow, so that we can solve exactly 
the continuity equation for the density. 
This is a great advantage, since we can tackle with the dynamics 
at the nonlinear stage. 
Next, we compare the Newtonian treatment with the general relativistic one. 
The Newtonian treatment is often used as a good approximation 
for the region $l/L \ll 1$, where $l$ is the length scale of fluctuations 
of fluids and $L$ corresponds to the Hubble radius.   
Thus, the treatment is restricted within small scales, though it
enables us to understand its results quite intuitively. 
On the other hand, there are no restrictions on scales 
in the general relativistic treatment. 
Hence, the relativistic version of the Lagrangian description is 
most useful for studying highly nonlinear region in the expanding 
universe up to the caustic formation. 
For dust, the Lagrangian description is formulated generally 
\cite{Kasai,RMKB,AK}. 
Now, we extend it to the case of fluid with pressure. 
This case include the radiation dominated era in the early
universe and collapsing region in which the velocity dispersion 
grows significantly at the late stage. 

A key idea in the Lagrangian description of the universe is 
as follows \cite{Zeldovich,Buchert89}: 
We illustrate it in the Newtonian cosmology for simplicity. 
The density is {\it exactly} obtained along the fluid flow. 
The Poisson equation for the cosmological Newtonian potential $\phi$ is 
\begin{equation}
\triangle\phi=4\pi G (\rho-\rho_b) , 
\end{equation}
where $\rho_b$ denotes a density in a background universe. 
This is estimated as 
\begin{equation}
\Bigl( \frac{L}{l} \Bigr)^2 \phi \sim \delta , 
\end{equation}
where $\delta$ denotes the density contrast. 
In the cosmological situation, $\phi$ can be safely considered as
small, even if the density contrast blows up in the small scale. 
Actually, this occurs when $l/L$ goes to zero, 
namely in the small scale. 
We can expect that the idea works well also in the relativistic case, 
if $\phi$ is replaced by the metric. 
The fluid flow approach has been discussed by several authors 
\cite{Hawking,Ellis71,Olson,LM}. 
However, their treatment is not satisfactory, since their formulations 
are based on the fluid flow but they solve their equations 
perturbatively by splitting the flow into the background and 
perturbed parts. 
Hence, their approach is still restricted within the small density
contrast. 

In section 2, we consider the perfect fluid. 
It is shown that the entropy and the vorticity are determined exactly 
based on the Lagrange condition. 
Under this condition, section 3 presents a perturbative 
Lagrangian approach.  
Conclusion is given in section 4. 
Greek indices run from $0$ to $3$, and Latin indices from $1$ to $3$. 
We use the unit, $c = 1$.

\section{Lagrangian description} 
Let us consider a universe filled with the perfect fluid, whose 
energy momentum tensor is written as 
\begin{equation}
T^{\mu\nu}=(\varepsilon+P) u^{\mu} u^{\nu} +P g^{\mu\nu} . 
\end{equation}
The conservation law becomes 
\begin{eqnarray}
(\varepsilon u^{\mu})_{;\mu}+P u^{\mu}{}_{;\mu}&=&0 , \\
(\varepsilon+P)u_{\mu;\nu}u^{\nu}+P_{,\nu}\gamma^{\nu}_{\mu}&=&0 , 
\end{eqnarray}
where we defined the projection tensor as 
\begin{equation}
\gamma^{\mu}_{\nu}=\delta^{\mu}_{\nu}+u^{\mu}u_{\nu} . 
\end{equation}

Here, we assume the barotropic equation of state as $P=P(\varepsilon)$, 
so that we can introduce the entropy and the enthalpy respectively as 
\cite{Ehlers,Weinberg}
\begin{eqnarray}
s&=&\exp{\int\frac{d\varepsilon}{\varepsilon+P}} , \\
h&=&\exp{\int\frac{dP}{\varepsilon+P}} . 
\end{eqnarray}
Then, the conservation law is re-expressed as 
\begin{eqnarray}
(s u^{\mu})_{;\mu}=0 , 
\label{entropyeq}\\
u_{\mu;\nu}u^{\nu}+(\ln{h})_{,\nu}\gamma^{\nu}_{\mu}=0 . 
\label{eom}
\end{eqnarray}

Furthermore, we define the vorticity as 
\begin{eqnarray}
\omega^{\mu}=\frac12 \epsilon^{\mu\alpha\beta\gamma}u_{\alpha} 
u_{\beta;\gamma} , 
\end{eqnarray}
where $\epsilon^{\mu\alpha\beta\gamma}$ denotes the complete
anti-symmetric tensor with 
$\epsilon^{0123} = 1/\sqrt{- g}$ and $g\equiv\det(g_{\mu\nu})$. 
Then, the Beltrami's equation for the vorticity is written as 
\begin{equation}
\Bigl(\frac{h \omega^{\mu}}{s}\Bigr)_{;\nu}\frac{u^{\nu}}{h} 
=\Bigl(\frac{u^{\mu}}{h}\Bigr)_{;\nu}\frac{h \omega^{\nu}}{s} . 
\label{beltrami} 
\end{equation}

To this point the treatment is fully covariant. 
In the following, we introduce the Lagrangian coordinate, 
\begin{equation}
x^{\mu}=(\tau, x^i) , 
\end{equation}
where $\tau$ is the proper time and $x^i$ is constant 
along the fluid flow. 
This gives us the Lagrangian condition (e.g. \cite{Friedrich}), 
in which the matter 4-velocity take components of  
\begin{equation}
u^{\mu}=(1,0,0,0) . 
\label{lagrange}
\end{equation}
Under this condition, we have $g_{00}=-1$ and $u_{\mu} = (-1, g_{0i})$. 
In the Lagrangian description, the entropy conservation equation 
$(\ref{entropyeq})$ is simply 
\begin{equation}
(s \sqrt{-g})_{,0} = 0 . 
\label{entropyeq2}
\end{equation}
Therefore, 
\begin{equation}
s (\bbox{x},t) = 
 \sqrt{ \frac{g(\bbox{x},t_0)}{g(\bbox{x},t)} } s (\bbox{x},t_0) .
\label{entropycons}
\end{equation}
The relativistic Beltrami equation (\ref{beltrami}) also becomes
simply
\begin{equation} 
\Bigl( \frac{h \omega^i}{s} \Bigr){}_{,0}=0 , 
\end{equation}
which is integrated to give 
\begin{equation}
\frac{h \omega^i}{s} =  
\left.\frac{h \omega^i}{s}\right|_{t_0} . 
\label{cauchy} 
\end{equation}
This is also expressed as 
\begin{equation}
h \omega^i(\bbox{x},t) = 
 \sqrt{ \frac{g(\bbox{x},t_0)}{g(\bbox{x},t)} } 
 h \omega^i(\bbox{x},t_0) .
\label{cauchy2}
\end{equation}
The $\omega^0$ component is not independent of $\omega^i$: From  
$ \omega^{\mu} u_{\mu}=0 $,  
we obtain 
\begin{equation}
\omega^0=g_{0i} \omega^i . 
\end{equation}

The result Eq. (\ref{cauchy}) tells us that the vorticity is coupled
to the entropy enhancement and vice versa. In particular, if the vorticity 
does not vanish exactly at an initial time, the vorticity will blow up 
as the entropy grows larger and larger (i.e. in the collapsing region), 
even if it has only the decaying mode in the linear perturbation theory. 
It should also be emphasized that our results Eqs. (\ref{entropycons}) 
and (\ref{cauchy}) in the fully general relativistic treatment 
precisely correspond to those in the Newtonian case. 

\section{Perturbative approach} 
In the previous section, we solved exactly the equations for the
entropy and the vorticity. The results Eqs. (\ref{entropycons}) and
(\ref{cauchy2}) show that $s$ and $\omega^i$ 
are completely written in terms of the determinant of the metric
tensor and their initial values. 
Here, we shall obtain perturbatively the metric at the linear order. 

The Einstein equation is decomposed with respect to the fluid flow: 
\begin{eqnarray}
G_{\mu\nu}u^{\mu}u^{\nu}&=&8\pi G \varepsilon , 
\label{H} \\
G_{\mu\nu}u^{\mu}\gamma^{\nu}_{\ \alpha}&=&0 , 
\label{M} \\
G_{\mu\nu} \gamma^{\mu}_{\ \alpha} \gamma^{\nu}_{\ \beta}&=&
8\pi G P \gamma_{\alpha\beta} . 
\label{E} 
\end{eqnarray}
The Euler equation $(\ref{eom})$ is rewritten as 
\begin{equation}
(h u_i)_{,0}+h_{,i}=0 . 
\label{euler}
\end{equation}

We assume that the background is spatially flat
Friedmann-Lema\^itre-Robertson-Walker (FLRW) universe. 
The extension to the spatially non-flat case must be a straightforward task. 
The perturbed metric is decomposed into 
\begin{eqnarray}\label{permet}
g_{0i}&=& B_{,i}(\bbox{x}) + b_{i}(\bbox{x}) , \\
g_{ij}&=& a^2 \Bigl(\delta_{ij} + 
         2 H_L \delta_{ij} + 2H_{T}{}_{,ij}+
          (h_{i,j}+h_{j,i}) + 2 H_{ij} \Bigr) , \nonumber
\end{eqnarray}
where $B$, $H_L$, and $H_T$ are scalar mode quantities, 
$b_i$ and $h_i$ are the vector (transverse) mode, and $H_{ij}$
is the tensor (transverse-traceless) mode satisfying 
\begin{eqnarray}
b^i_{\ ,i}&=&0 , \\
h^i_{\ ,i}&=&0 , \\
H^i_{\ i}&=&0 , \\
H^{ij}_{\ \ ,j} &=&0 .  
\end{eqnarray}
Raising and lowering indices of the perturbed quantities are done 
by $\delta^{ij}$ and $\delta_{ij}$. 

\subsection{Residual gauge freedom in the Lagrange condition}

The general gauge transformation to the first order is induced 
by the infinitesimal coordinate transformation 
\begin{equation}
\tilde{x}^{\mu}=x^{\mu}+\xi^{\mu} . 
\end{equation}
The changes due to the gauge transformation are
\begin{eqnarray}
\delta_{\xi} g_{\mu\nu} 
&=& -g_{\mu\nu,\alpha}\xi^{\alpha}-g_{\mu\alpha}\xi^{\alpha}{}_{,\nu}
-g_{\nu\alpha}\xi^{\alpha}{}_{,\mu} , 
\label{deltag} \\
\delta_{\xi}u^{\mu} 
&=&\xi^{\mu}{}_{,\nu}u^{\nu}-u^{\mu}{}_{,\nu}\xi^{\nu} . 
\label{deltau}
\end{eqnarray}
In order to leave the Lagrangian condition unchanged, we have 
$\delta_{\xi}u^{\mu} = 0$, which leads to $\xi^{\mu}{}_{,0}= 0$. 
Therefore, 
\begin{equation}
\xi^{\mu}=\xi^{\mu}(\bbox{x}) . 
\label{gaugeTL} 
\end{equation}

For the spatially flat background, $\xi^{\mu}$ are decomposed into each 
mode, 
\begin{equation}
\xi^{\mu}(\bbox{x})=(T,\delta^{ij} L_{,j}+\ell^i) , 
\end{equation}
where the vector mode quantity $\ell^i$ satisfies $\ell^i_{\, ,i} = 0$. 
Hence we obtain 
\begin{eqnarray}
\tilde{B} &=& B + T(\bbox{x}), \label{gaugeB}\\
\tilde{H}_L &=& H_L -  \frac{\dot{a}}{a} T(\bbox{x}), \label{gaugeHL}\\
\tilde{H}_T &=& H_T - L(\bbox{x}), \label{gaugeHT}\\
\tilde{h}_i &=& h_i - \ell_i(\bbox{x}) ,  \label{gaugehi} 
\end{eqnarray}
where an overdot ($\dot{}$) denotes $\partial/\partial t$. 
The vector mode quantity $b_i$ and the tensor mode quantity
$H_{ij}$ are gauge-invariant. 

For a general case of $P=P(\varepsilon)$, the scale factor takes a 
complicated form. 
For simplicity, let us take the equation of state as 
\begin{equation}
P=\frac13\varepsilon . 
\end{equation}

\subsection{Scalar perturbations}

The Einstein equations for the scalar perturbations are
\begin{eqnarray}
\Bigl( \frac{\dot{a}}{a} \Bigr)^{\bbox{\cdot}} B+\dot{H}_L &=& 0 , 
\label{eqscalar1} \\
\ddot{H}_L + 4 \frac{\dot{a}}{a} \dot{H}_L - \frac23\frac{\dot{a}}{a} 
\nabla^2 H_{L}+\frac13\frac{\dot{a}}{a} \nabla^2 \dot{H}_{T}
-\frac13 \frac{\dot{a}}{a} \nabla^2 B  
&=& 0, 
\label{eqscalar2}  \\
\ddot{H}_T + 3 \frac{\dot{a}}{a} \dot{H}_T &=& \frac{1}{a^2} 
\Bigl( H_L + \frac{1}{a}{(a B)}^{\bbox{\cdot}} \Bigr) , 
\label{eqscalar3}
\end{eqnarray}
where $\nabla^2=\delta^{ij}\partial_i \partial_j$. 
We introduce the conformal time $\eta$ as 
\begin{equation}
d\eta=\frac{dt}{a} . 
\end{equation}

We expand all quantities in Fourier's series; for instance 
\begin{equation}
\Psi=\int d^3k \Psi_{\bbox{k}}(\eta) \exp{(i\bbox{k}\cdot\bbox{x})} , 
\end{equation}
where the subscript $\bbox{k}$ denotes the Fourier coefficient. 
We define $\theta$ as 
\begin{equation}
\theta=\frac{k\eta}{\sqrt{3}} , 
\end{equation}
where $k=|\bbox{k}|$. 
We are in position to solve Eqs. (\ref{eqscalar1})-(\ref{eqscalar3}) 
and obtain 
\begin{eqnarray}
B_{\bbox{k}} &=& \frac14 \Bigl( \frac{\sqrt{3}}{k} \Bigr)^2 
\Bigl[ \Psi_{\bbox{k}}(-2+\cos\theta+\theta\sin\theta) 
+\Phi_{\bbox{k}}(-2\sin\theta+\theta\cos\theta) \Bigr] , \\
H_{L \bbox{k}} &=& \Psi_{\bbox{k}} \theta^{-2}(1-\cos\theta) 
+\Phi_{\bbox{k}} \theta^{-2}\sin\theta , \\
H_{T \bbox{k}} &=& \Bigl( \frac{\sqrt{3}}{k} \Bigr)^2 
\Bigl[ \Psi_{\bbox{k}} \Bigl( \theta^{-1}\sin\theta-\frac12\cos\theta 
-\frac12\Bigr) 
+\Phi_{\bbox{k}} \Bigl(\theta^{-1}\cos\theta+\frac12\sin\theta 
-\theta_0^{-1}\Bigr) \Bigr] , 
\end{eqnarray}
where $\theta_0$ denotes $\theta$ at the initial time and 
we used the residual gauge freedom to set initially $H_{T \bbox{k}}=0$. 

The initial entropy field $s(\bbox{x}, t_0)$ is also expressed by
the metric. When the initial entropy contrast $\delta \equiv
(s-s_b)/s_b$ is sufficiently small, we obtain from
Eq. (\ref{H})
\begin{eqnarray}
\delta_{\bbox{k}}(\eta_0) = \frac43 \Bigl[ && 
\Psi_{\bbox{k}} \Bigl( 3\theta_0^{-2} (\cos\theta_0 -1) 
+3\theta_0^{-1} \sin\theta_0 -\frac32\cos\theta_0 \Bigr) \nonumber\\ 
&&+\Phi_{\bbox{k}} \Bigl( -3\theta_0^{-2} \sin\theta_0 
+3\theta_0^{-1} \cos\theta_0 +\frac32\sin\theta_0 \Bigr) \Bigr] . 
\end{eqnarray}

\subsection{Vector perturbations}
The Einstein equations for the vector perturbations are 
\begin{eqnarray}
\nabla^2 \Bigl( \dot{h}_i - \frac{1}{a^2}b_i\Bigr) &=& 
4 \Bigl(\frac{\dot{a}}{a}\Bigr)^{\bbox{\cdot}} b_i ,
\label{eqvector1}\\
\Bigl( a^3 \dot{h}_i - a b_i \Bigr)^{\bbox{\cdot}} &=& 0 . 
\label{eqvector2}
\end{eqnarray}
Introducing $\beta_i$ as 
\begin{equation}
b_i(\bbox{x})= t^{1/2} \nabla^2 \beta_i(\bbox{x}) , 
\end{equation}
Eq. (\ref{eqvector1}) is solved to give 
\begin{equation}
h_i=2 (t-t_0)^{1/2} \nabla^2 \beta_i(\bbox{x}) 
+ 4 (t-t_0)^{-1/2} \beta_i(\bbox{x}) ,
\label{vector} 
\end{equation} 
where we again used the residual gauge freedom
(cf. Eq. (\ref{gaugehi})) to set
$h_i(\bbox{x},t_0) = 0$. 

\subsection{Tensor perturbations}

The equation for the tensor perturbations is 
\begin{equation}
\ddot{H}_{ij} + 3\frac{\dot{a}}{a}\dot{H}_{ij}
-\frac{1}{a^2}\nabla^2 H_{ij} = 0 . 
\end{equation}
This is a homogeneous wave equation in the expanding universe. 
The solutions are well known and we will not discuss the detail here. 
See, e.g., \cite{Weinberg}. 

\subsection{Discussion}

As for the metric, our result is identical to that of the linear 
perturbation theory (e.g. \cite{Bardeen}). 
However, it should be emphasized our Lagrangian approach does not 
rely on the assumption that {\it the entropy contrast should be small}. 
It is actually an important advantage that we can use (or extrapolate) 
the well-known solutions of the linear theory to express 
the non-linear behavior of the entropy.

It is worthwhile to mention that there are some choices of 
a temporal coordinate. 
Here, we have adopted the proper time, for simplicity. 
We have another choice 
$u^{\mu}=(h, \vec 0) $
\cite{Ellis73}. 
Then, Eq. (\ref{euler}) is solved as 
$g_{0i}={f_i(\bbox{x})}/{h^2}$. 
However, we find 
$g_{00}=-1/{h^2}$, 
which implies that it seems rather tedious to solve 
the Einstein equation.

\section{Conclusion}
The Lagrangian description in the relativistic cosmology has been 
extended to the case of fluid with pressure. 
It is applicable to even highly nonlinear regions up to the caustic 
formation, since the entropy and the vorticity are obtained exactly 
along the fluid flow. 
In this approach, dynamical variables are the metric but not material 
variables thanks to the Lagrange condition.
The present description includes the dust cosmology as a special 
case $(P=0)$ \cite{AK}. 

As future subjects along this Lagrangian approach, 
it would be interesting to study the primordial black hole cosmology 
(e.g. \cite{Carr}) and the averaging problem 
in inhomogeneous cosmologies (e.g. \cite{Kasai,Futamase,Buchert99}). 
Furthermore, self-interacting cold dark matter (field) 
models have been recently discussed as a way to alleviate 
inconsistencies between the standard cold dark matter 
scenario and the current observation on galactic and 
subgalactic scales ($\leq$ few Mpc) \cite{PV,SS,Goodman}. 
It may be also an important application of the present formalism 
to study such a nonlinear scalar dynamics, which is modeled 
by fluids \cite{Madsen}.

\section*{Acknowledgment}
The author would like to thank M. Kasai, T. Buchert, S. Matarrese, 
N. Sugiyama and K. Yamamoto for fruitful discussion. 
He also would like to thank Gerhard B\"orner for hospitality at
Max-Planck-Institut f\"ur Astrophysik, where a part of this work 
was done. 
This work was supported in part by a Japanese Grant-in-Aid 
for Scientific Research from the Ministry of Education, Science 
and Culture, No. 11740130.

\end{document}